%% file: main.tex
\documentclass[a4paper]{article}

\usepackage{INTERSPEECH2021}
\usepackage{multirow}
\title{SA-SASV: An End-to-End Spoof-Aggregated Spoofing-Aware Speaker Verification System}

\name{Zhongwei Teng$^1$, Quchen Fu$^1$, Jules White$^1$, Maria E. Powell$^2$, Douglas C. Schmidt$^1$}
\address{
  $^1$Dept. of Computer Science, Vanderbilt University
  \\
  $^2$Dept. of Otolaryngology--Head and Neck Surgery, Vanderbilt University Medical Center
  }
\email{}

\begin{document}

\maketitle
\begin{abstract}
Research in the past several years has boosted the performance of automatic speaker verification systems and countermeasure systems to deliver low Equal Error Rates (EERs) on each system. However, research on joint optimization of both systems is still limited. The Spoofing-Aware Speaker Verification (SASV) 2022 challenge was proposed to encourage the development of integrated SASV systems with new metrics to evaluate joint model performance. This paper proposes an ensemble-free end-to-end solution, known as Spoof-Aggregated-SASV (SA-SASV) to build a SASV system with multi-task classifiers, which are optimized by multiple losses and has more flexible requirements in training set. The proposed system is trained on the ASVSpoof 2019 LA dataset, a spoof verification dataset with small number of bonafide speakers. Results of
SASV-EER indicate that the model performance can be further improved by training in complete automatic speaker verification and countermeasure datasets.

\end{abstract}
\input{body}

\bibliographystyle{IEEEtran} 
\bibliography{main}

\end{document}

%% file: body.tex
\vspace{-0.2cm}
\section{Introduction}

Automatic speaker verification (ASV) systems have shown the ability to provide biometric authentication of users for applications that require robust reliability in changing acoustic environments, including resistance to malicious attacks~\cite{campbell2006support,dehak2010front,heigold2016end,snyder2018x,desplanques2020ecapa}.
However, current ASV systems are still vulnerable to spoofing attacks, such as text-to-speech (TTS)~\cite{oord2016wavenet,morise2016world,schroder2011open} and voice conversion (VC)~\cite{matrouf2006effect}. ASV systems can also be deceived and manipulated by malicious entities using generated speech.

To overcome bottlenecks in spoofing and countermeasure research for ASVs, a series of  ASVSpoof challenges have been proposed since 2015 to help encourage the development of robust countermeasure (CM) systems~\cite{wu2015asvspoof, kinnunen2017asvspoof,todisco2019asvspoof,yamagishi2021asvspoof}, which can complement ASV systems with an anti-spoof model. The anti-spoof model  provides a "spoof confidence" score to help filter out spoofing attacks. 
Metrics on the ASVSpoof challenge are based on the minimum tandem detection cost function (t-DCF)~\cite{kinnunen2020tandem}, which can evaluate the performance of CM systems on fixed ASV systems with pre-determined output scores. Rather than developing CM and ASV systems independently, a neglected research question is whether we can develop an integrated system where CM and ASV system can be optimized together, so that a single verification score is able to determine whether an input speech sample is a target speaker, while also accounting for potential spoofing attacks.

To encourage research on integrated Spoofing-Aware Speaker Verification (SASV) systems, the SASV Challenge 2022~\cite{jung2022sasv} was proposed using the ASVSpoof 2019 Logical Access Dataset with new metrics, SASV-EER.
In the challenge, a single score determines if the input speech sample is the target speaker. Non-target inputs include zero-effort and spoofed impostors. The SASV challenge provides two baseline systems by applying different fusion strategies (score-level fusion and embedding-level fusion) to pre-trained ASV and CM systems. 

\begin{figure}[hptb]
    \centering
    \includegraphics[width=5.2cm,height=6.5cm]{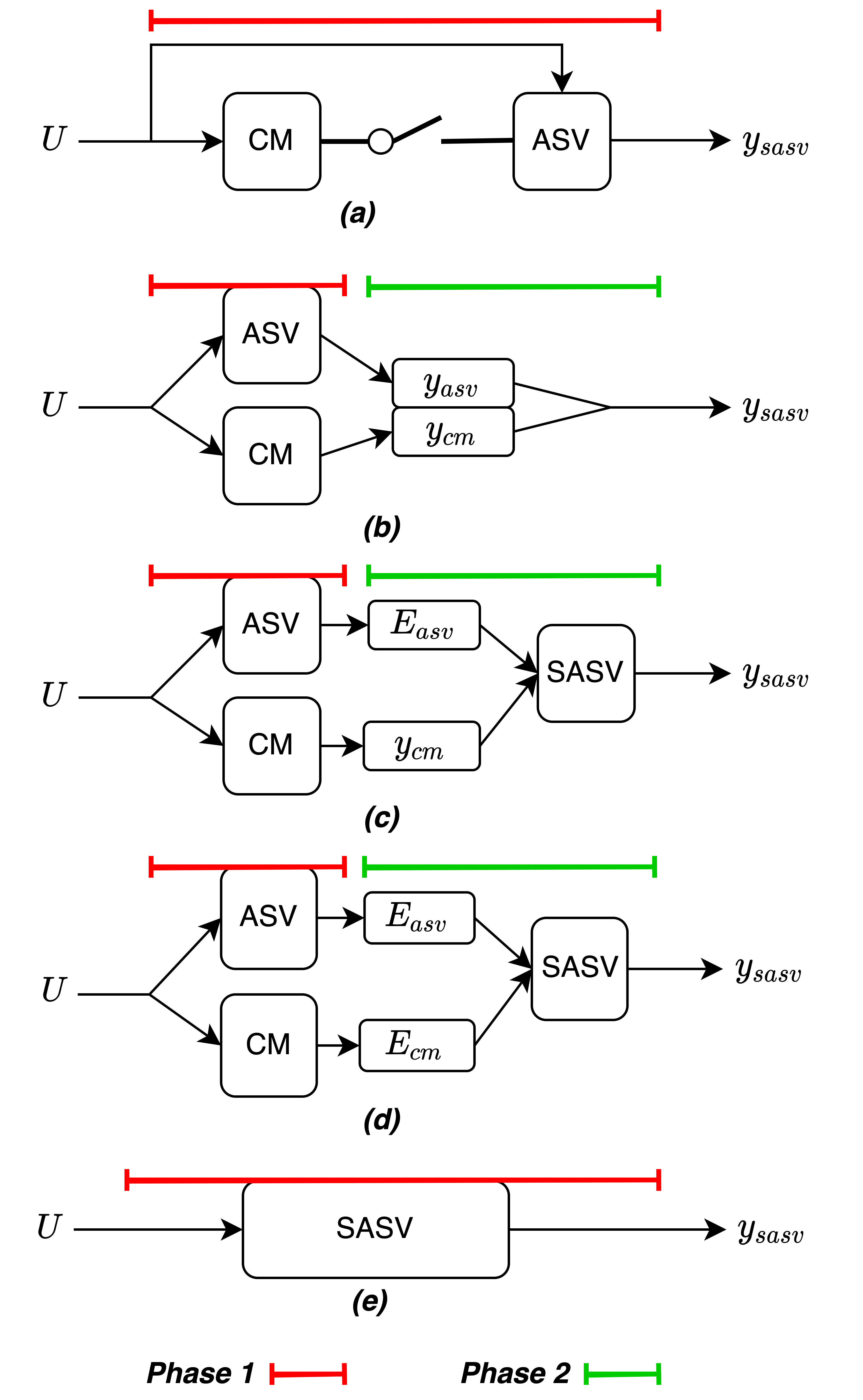}
    \vspace{-0.2cm}
    \caption{Feasible Solutions to Build Integrated SASV Systems. }
    \label{fig.solutions}
        \vspace{-0.3cm}
\end{figure}

Figure~\ref{fig.solutions} shows potential solutions to the SASV problem.  Red/green lines indicate the following training stages: (a) Cascaded ASV/CM systems, (b) Fusions of scoring prediction, (c)Fusions of scoring and feature embedding, (d)Fusions of feature embedding, and (e)End-to-End SASV systems.

This paper proposes a fully trainable end-to-end SASV system, called Spoof-Aggregated Spoofing Aware Speaker Verification System (SA-SASV), that combines a pre-trained ASV system with a lightweight raw waveform encoder to form the overall encoder~\cite{teng2021complementing}. This paper expands upon our prior experience that showed how encoding can be a key aspect of these types of anomaly detection problems~\cite{teng2021complementing,pan2019detecting,fu2021fastaudio}. Multiple classifiers and spoof-source-based triplet loss functions are employed to enhance model performance in generating the shared SASV feature space.  

The remainder of the paper is organized as follows: Section~\ref{sec.related} reviews related research on SASV systems; Section~\ref{sec.model} discusses the model architecture of our SA-ASAV Systems; Section~\ref{sec.results} analyzes experiment results; and Section~\ref{sec.conclusion} presents concluding remarks.

\vspace{-0.2cm}
\section{Related Work}
\label{sec.related}
The SASV system aims to build a single system to reject utterances from zero-effort and spoofed speech. Previous work focused on two solutions to this problem: ensemble SASV solutions and integrated single system solutions.

Ensemble SASV solutions take fixed outputs from pre-trained ASV and CM systems and apply varying fusion strategies to generate a single SASV score for both tasks. Sizov et al.~\cite{sizov2015joint} was the first to apply i-vectors and a PLDA back-end for joint modeling of speaker verification and spoof detection. At the score level, Todisco et al.~\cite{todisco2018integrated} proposed a two-dimensional score modeling method to get a single score threshold for both ASV and CM systems.

Shim et al.~\cite{shim2020integrated} discusses a back-End modular approach to train embeddings from pre-trained fixed ASV systems and spoofing predictions from CM systems to predict final SASV scores.
In addition to scoring ensembles, fusions based on embeddings from different models have also been tested. For example,
Gomez-Alanis et al.~\cite{gomez2020joint} proposed DNN-based integration methods to train three types of embeddings from ASV and CM systems jointly.

The target task of an integrated single SASV system is to build an end-to-end system that simultaneously classifies speech based on whether or not it is from the target speaker and is authentic non-spoofed speech. 
Zhao et al.~\cite{zhao2022multi} built an SR-ASV system with two classifiers to get CM scores and ASV scores from shared layers and the final decision is based on both the CM and ASV scores. Li et al.~~\cite{li2020joint} applied speaker-based triplet loss to train multi-task classification networks to make a joint decision on anti-spoofing and ASV.

As a form of integrated single SASV system, our method explores the shared feature space of SASV tasks. To obtain proper embeddings for speech from the multiple encoders that we employ, both hand-crafted features and raw waveforms are input into SA-SASV. We first discuss the feasibility of optimizing the SASV feature space by aggregating spoofed voice samples based on their spoofing sources. The proposed model was trained with multiple loss functions, including spoof source-based triplet loss. The final decision by our model is based on cosine similarity and CM scores from same model. 

\vspace{-0.2cm}
\section{AS-DGASV Model Architecture}
\label{sec.model}
\vspace{-0.1cm}

\begin{figure}[tbhp]
    \centering
    \includegraphics[width=6cm,height=3cm]{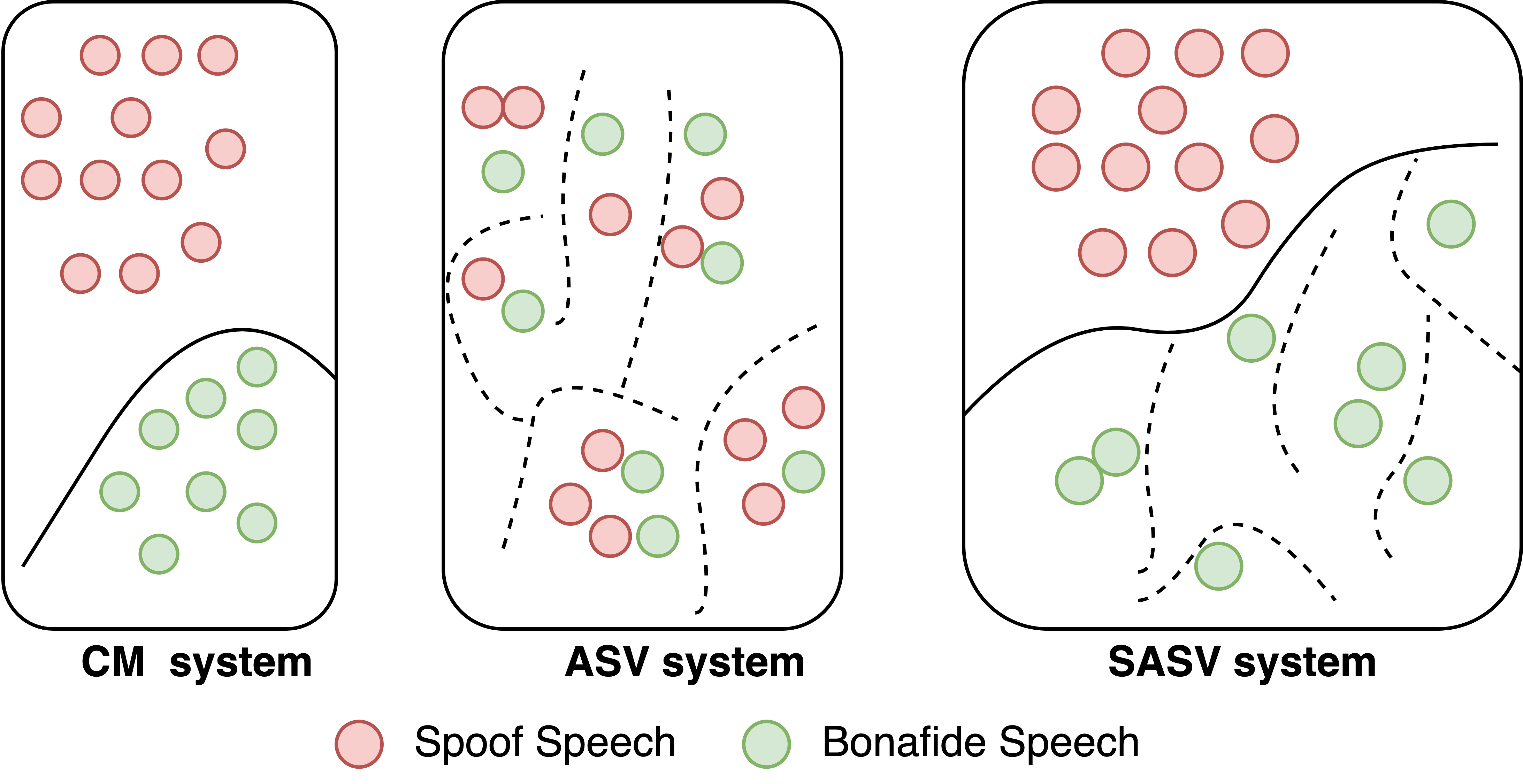}
    \vspace{-0.2cm}
    \caption{Desired Speech Sample Classification Distribution of Different Spoof Detection Systems.}
    \label{fig.dist}
    \vspace{-0.2cm}
\end{figure}

Compared to independent CM and ASV models, the ideal feature space learned from SASV models should have the following characteristics:
(1) spoofed and bonafide speech should be densely aggregated so that obvious margins can be drawn to separate them and (2) in the clusters of bonafide speech sources from different speakers should be sparsely distributed so that models can distinguish between different speakers. Figure~\ref{fig.dist} shows how the SASV system integrates the CM and ASV systems so that there are two types of boundaries to separate spoof/bonafide speech and target/non-target speakers. 

\begin{figure}[hbpt]
    \centering
    \includegraphics[width=8cm,height=4cm]{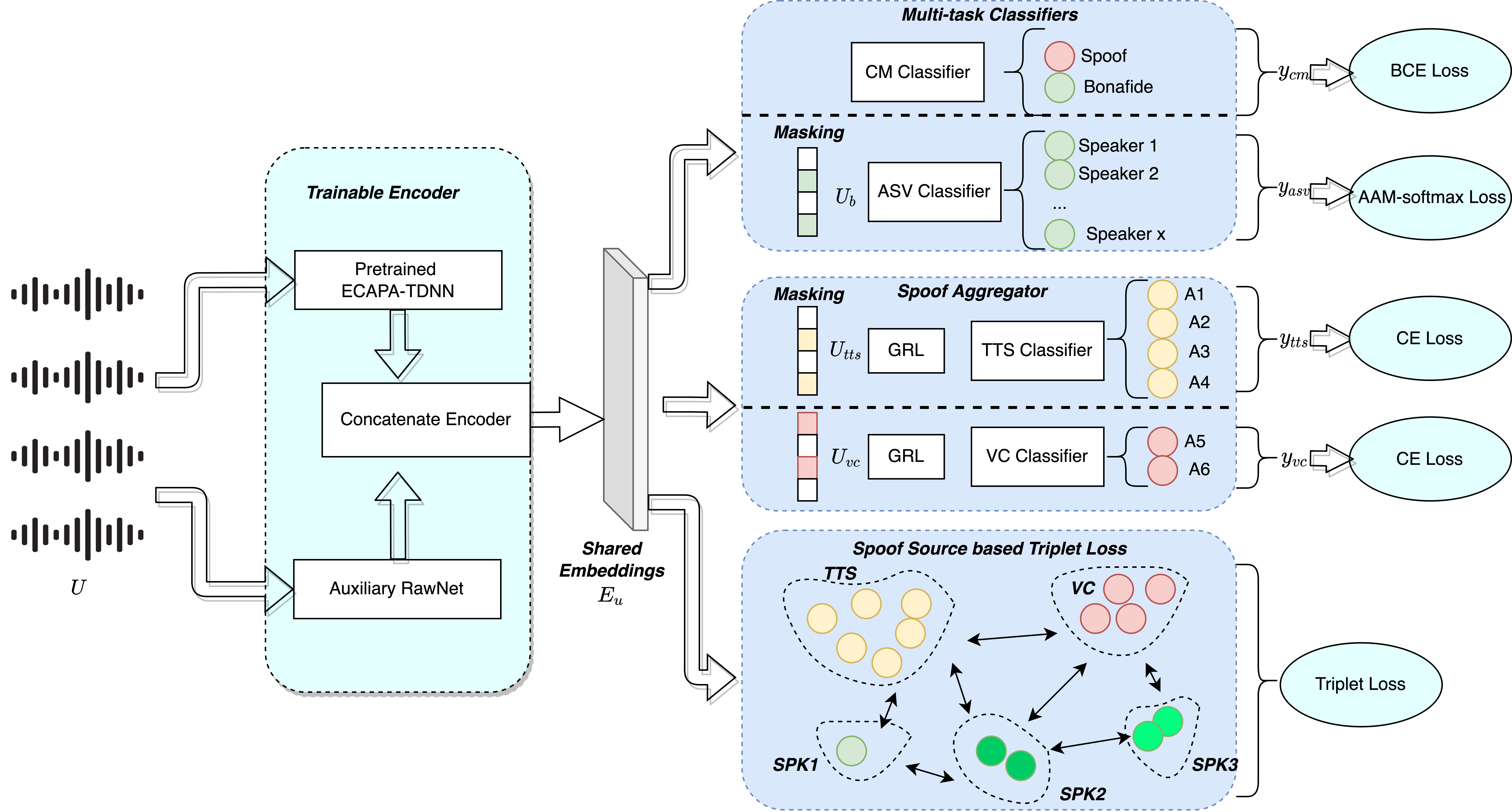}
    \vspace{-0.3cm}
    \caption{Model Structure of the SA-SASV System.}
    \label{fig.structure}
        \vspace{-0.4cm}
\end{figure}

To achieve optimized feature space in a SASV system, we proposed the SA-SASV model, whose decode consists of three parts: multi-task classifiers, spoof aggregators, and spoof-source-based triplet loss, as shown in Figure~\ref{fig.structure}. 
This figure shows how shared embedding is fed into multiple classifiers and how the feature space from the encoders is optimized by combinations of various loss functions. This fully trainable model takes both raw waveforms and hand-crafted features as input and multiple losses are used to optimize feature embedding.

\vspace{-0.2cm}
\subsection{The ARawNet Encoder}
Previous research shows that the best-performing ASV systems~\cite{desplanques2020ecapa} and CM systems~\cite{jung2021aasist}, take hand-crafted features and raw waveforms, respectively, indicating distinctive features among each type of input that are useful for identifying speakers and spoofing attacks. It is hard, however, to simply merge existing state-of-the-art ASV and CM systems together to develop an end-to-end model, due to the resulting large model size and high computational complexity. We use our previously published ARawNet architecture~\cite{teng2021complementing} to help overcome this limitation. Our encoder combines a pre-trained ASV system (ECAPA-TDNN) and a lightweight raw waveform encoder to enable simultaneous analysis of both learned features and raw wave forms.

We denote input utterance as $U$.  An utterance's embedding, $E_u$, can be described as shown in Equation~\ref{eq-emb}, where $F_{asv}$ is a pre-trained ECAPA-TDNN, $F_{raw}$ is an un-trained auxiliary raw encoder, and $F_c$ is a concatenating encoder that handles outputs from $F_{asv}$ and $F_{raw}$.

\vspace{-0.2cm}
\begin{equation}
\label{eq-emb}
    E_u = F_{c}(F_{asv}(U), F_{raw}(U))
\end{equation}
\vspace{-0.1cm}

\vspace{-0.6cm}
\subsection{Multi-task Classifiers}

Since end-to-end SASV systems need to determine if input speech is bonafide---and if so, if it is the target speaker---this problem is formulated as a multi-task classification problem. Two classifiers are used to predict spoof attacks and speaker id independently, with shared feature embeddings from the encoder. The CM classifier $C_{cm}$ receives all inputs and predicts confidence scores, indicating if the input is believed to represent a spoofing attack. A bonafide mask layer is placed before the ASV classifier, $C_{asv}$, so that losses produced by the ASV classifier are only from bonafide speech. Binary cross entropy(BCE) loss and AAM-softmax loss are used for the CM and ASV classifiers.
% as described in   Equation~\ref{equ.bce} and Equation~\ref{equ.aam}\cite{xiang2019margin} respectively.

% \begin{equation}
% \label{equ.aam}
%     L_{asv} = - \frac{1}{N} \Sigma^{N}_{i=1} log \frac{e^{s(cos(\theta_{y_i,i})+m )}}{e^{s(cos(\theta_{y_i,i})+m)}+ \Sigma^{j}_{j=1,j\neq i}e^{s(cos(\theta_{j,i}))}}
% \end{equation}

% \begin{equation}
% \label{equ.bce}
%     L_{cm} = - \frac{1}{N} \Sigma^{N}_{i=1} y_{i}^{cm}  logC_{cm}(E) + (1-y_{i}^{cm})log(1-C_{cm}(E))
% \end{equation}

\subsection{Spoof Aggregator}
\begin{figure}[htpb]
    \centering
    \includegraphics[width=3.5cm,height=3cm]{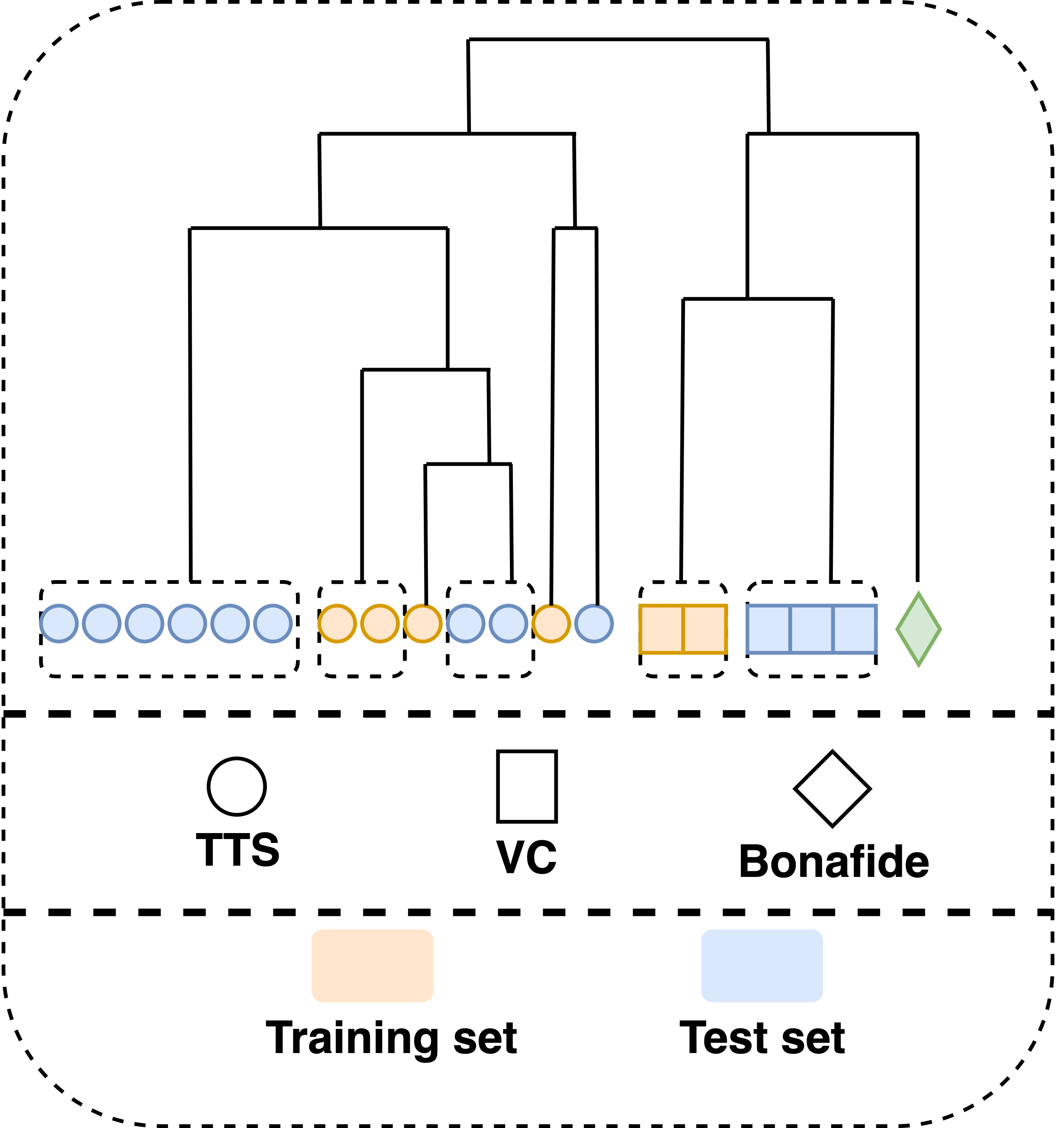}
      \vspace{-0.1cm}
    \caption{Results of Agglomerative Clustering on the ASVSpoof 2019 LA Dataset.}
    \label{fig.spoof_agg}
    \vspace{-0.2cm}
\end{figure}

In the SASV task, utterances, $U$, consists of spoof attack samples, $U_s$, and bonafide speech samples, $U_b$.
As shown in Figure~\ref{fig.dist},  $U_s$ should have a relatively dense distribution in the shared feature space. It is hard, however, to aggregate the various spoofing attacks together due to their intrinsic differences in speech generation methods. This inherent difficulty in separating the two is evidenced by analyzing $U_s$ from different sources using agglomerative clustering~\cite{wang2020asvspoof}, as shown in Figure~\ref{fig.spoof_agg}.
These results indicate that $U_{tts}$ (which represents produced with Text-to-Speech(TTS)) and $U_{vc}$ (which represents samples from Voice Conversion(VC)) $U_{vc}$, tend to be closer in corresponding feature space.
We therefore conjecture that $U_{tts}$ and $U_{vc}$ should be aggregated into two clusters in the feature space of SASV systems.

We use two 
adversarial learning layers to construct a spoof aggregator so that $U_{tts}$ and $U_{vc}$ aggregate separately. We labeled the $U_{s}$ as $A1 \dots A6$, representing the spoof type, where $A1$ to $A4$ are from $U_{tts}$  and $A5$ to $A6$ are from $U_{vc}$. Followed by a masking layer, $E_{tts}$ and $E_{vc}$ are sent to $C_{tts}$ and $C_{vc}$, where each independently attempts to predict what spoof type $U_s$ corresponds to.
% The cross entropy loss for both classifiers is shown in Equation~\ref{equ.ce}
% \begin{equation}
% \label{equ.ce}
%     L_{tts} = L_{vc} = \frac{1}{N} \Sigma^{N}_{i}y_i^{A} log C_{spoof}(E)
% \end{equation}

Since we want our embedding, $E$, to mix spoof attacks from the same types of generation mechanisms together, so that $C_{tts}$ and $C_{vc}$ fail to distinguish different spoofing attack types, a gradient reverse layer(GRL) is added before the classifiers to maximize $L_{tts}$ and $L_{vc}$.

\vspace{-0.2cm}
\subsection{Spoof source based triplet loss}

The shared feature space from SASV systems tends to be differentiated by $U_{tts}$, $U_{vc}$, and different speakers $U_{spk_i}$.
In other words, the goal is for $E$ with the same labels to be relatively compactly clustered and the overall cluster separated from $E$ samples with different labels.
Boundaries between the $E$ samples with different labels should be distinct.
To help achieve this outcome, rather than applying speaker-based  triplet loss, we applied spoof source-based Triplet loss. 
Conventional triplet loss is described as Equation~\ref{equ.trip}:
\begin{equation}
    \vspace{-0.2cm}
    \label{equ.trip}
    L_{t} = \lVert E^{a}-E^{p} \rVert -  \lVert E^{a}-E^{n}  + m \rVert
\end{equation}

\begin{figure}[tbhp]
    \centering
    \includegraphics[width=4.5cm,height=5cm]{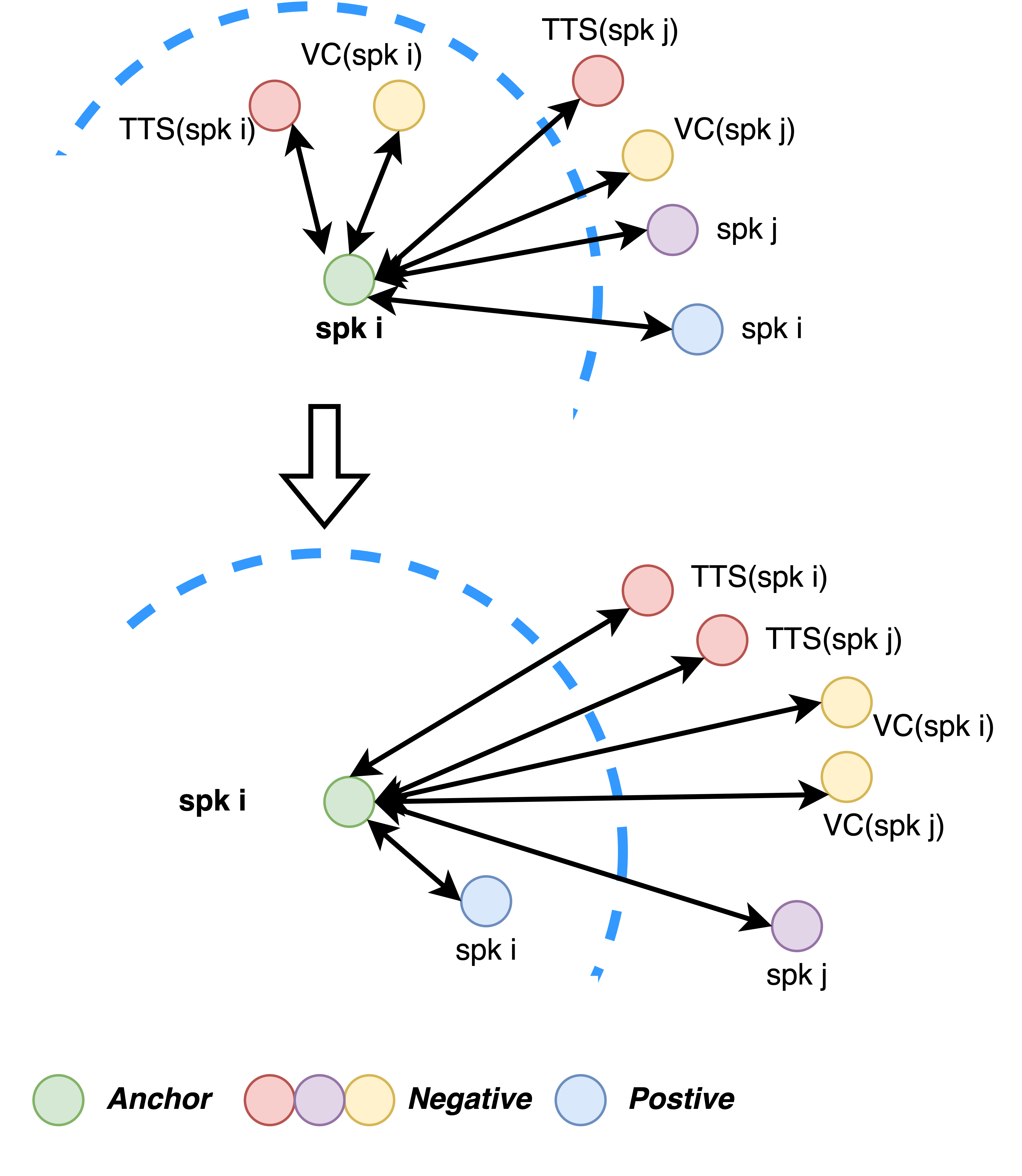}
        \vspace{-0.2cm}
    \caption{Training Based on Spoof-source Based Triplet Loss. }
    \label{fig.triplet}
    \vspace{-0.4cm}
\end{figure}

As shown in Figure~\ref{fig.structure}, $E_i$ is labeled as $TTS$, $VC$ and $SPK_i$, where $SPK_i$ indicates the $i th$ speaker. The goal is to cluster, $E_i$ samples, with same labels as densely as possible and scatter $SPK_i$ to make it far away from $SPK_j$ , $TTS$ and $VC$, as shown in Figure~\ref{fig.triplet}. 
This figure shows that positive samples (utterances with the same labels) are pulled closer and negative samples are pushed away.
Thus, for an utterance from speaker $i$, $U_{spk_i}$, the spoof source based triplet loss is updated as shown in Equation~\ref{equ.ssb}.
\vspace{-0.5cm}

\begin{equation}
\label{equ.ssb}
\begin{aligned}
    L_{st} = L_{t}(E_{a},E_{p}, E_{tts}) + L_{t}(E_{a},E_{p}, E_{vc}) \\ 
    +\Sigma^{N}_{j=0}L_{t}(E_{a},E_{p}, E_{spk_j}) i \neq j
\end{aligned}
\vspace{-0.3cm}
\end{equation}

\subsection{Overall Loss Function}
As shown in Figure~\ref{fig.structure}, the overall loss for  AS-DGASV is determined by all of its constituent decoders, which includes five different loss functions, as shown in Equation~\ref{equ.loss}. 
\vspace{-0.5cm}

\begin{equation}
\label{equ.loss}
    L_{sasasv} = L_{cm} + \lambda_1 L_{asv}+\lambda_2 L_{tts} +\lambda_3 L_{vc} + \lambda_4 L_{ts} 
\end{equation}

\vspace{-0.4cm}
\section{Analysis of Experimental Results}
\label{sec.results}
This section analyzes the results of experiments we conducted to. We analyzed our model performance with ablation study and compared with prior research in the SASV problem.

\vspace{-0.2cm}
\subsection{Experiment Setting}
\vspace{-0.2cm}
\input{tables/table_compare}
\textbf{Dataset.}
The SASV challenge permits the VoxCeleb2 dataset~\cite{chung2018voxceleb2} and the ASVspoof 2019 LA dataset~\cite{todisco2019asvspoof} for training the ASV and CM models.
The VoxCeleb2 database consists of over 1 million utterances from 6,112 speakers and is designed for the ASV task, without spoofed data. 
The ASVspoof 2019 LA dataset, on the other hand, is prepared for the CM tasks, containing 6 types of spoof attacks in the training set and another 11 types of spoof attacks in the evaluation set, where the SASV models are tested.
We use the VoxCeleb2 dataset to pre-train the ECAPA-TDNN and our model is fine-tuned on the ASVspoof 2019 LA dataset.
% As shown in Table~\ref{table.dataset}, the models need to generalize training attacks (A1-A6) to unseen attacks (A7-A19).

\textbf{Metrics.}
We evaluated our model performance based on the SASV-EER, which is the primary metric in the SASV challenge. 
% As shown in Table~\ref{table.metrics}, 
Only target speakers are labeled as positive and both non-target bonafide and spoof attacks are labeled as negative in the SASV-EER. The SV-EER and SPF-EER, are complements to SASV-EER, and reflect models' capability in different subsets of the full trials. 
Compared to the EER used in the ASVSpoof challenge, the SPF-EER only tests model performance in trials based on bonafide target speakers with spoofed speech.

% \begin{table}[hbt!]

% \centering
% \begin{tabular}{llll}
% \hline
%          & Target & Non-target & Spoof \\ \hline
% SV-EER   & +      & -          &       \\ \hline
% SPF-EER  & +      &            & -     \\ \hline
% SASV-EER & +      & -          & -     \\ \hline
% \end{tabular}
% \caption{Metrics to evaluate SASV systems.}
% \label{table.metrics}
% \end{table}

\textbf{Baseline.}
The SASV challenge provides two baseline models using  state-of-the-art ASV and CM systems with different fusion strategies.
\textbf{Baseline1} adopts a score-sum ensemble, which uses a naive sum function to integrate non-calibrated scores from the ASV and CM systems. 
While this method does not consider the difference between scores from different systems, scores of ASV systems are cosine similarity and scores of CM systems are from classifiers.
\textbf{Baseline2} uses an extra network as a fusion strategy that takes embeddings from pre-trained ASV and CM systems to produce the final scores.

\input{tables/table_results}

\vspace{-0.0cm}
\subsection{Results Discussion}
\vspace{-0.1cm}

\subsubsection{Ablation Study on the Proposed Model}
\vspace{-0.1cm}

\textbf{Configuration.} An ablation study was conducted to investigate the effects of the different components on the performance of the SA-SASV system. As shown in Table~\ref{table.results}, we evaluated our model with varying configurations of (1) just spoof source-based triplet loss, (2) just spoof aggregator, (3) and the two combined. Results indicate the absence of either component will reduce the SASV-EER of the SA-SASV model and configurations with all proposed sub-structures provide the best results on the ASVSpoof 2019 Dataset. By comparing SPF-EER, we can find spoof-source-based triplet loss boosts model performance in the countermeasure task in multi-task classification model. Our best results improve all three metrics and the SASV-ERR was improved from 8.75\% (baseline) to 4.86\%. 

\textbf{The proposed model shows different generalization capabilities in SV and SPF tasks.}
Even though the SV-EER of the model reaches 0 in the training stage, it has limited ability to generalize the SV task to the evaluation set, which only contains unseen speakers. As a result, the overall model performance drastically decreased due to SV-EER. We also noticed that, due to the overfitting problem, compared to SPF-EER, SV-EER in all models with different configurations tends to have unstable results. However, the SPF-EER of the model shows consistency from training to evaluation set, the best SPF-EER reaches 0.5, which is better than the baseline single CM system. 

In conclusion, the model can detect unseen spoof attacks and has trouble distinguishing unknown speakers in the evaluation set. We conjecture the performance difference stems from data distribution in the training set. Only 40 speakers are contained in the training set and the ASV task usually requires a larger number of speakers to build features of human utterance, e.g., 5,994 speakers are included in the VoxCeleb2 dataset.

Although parts of our encoder are pre-trained on the VoxCeleb2 dataset, it only gave our model a feasible initializing strategy. During the training stage, the bonafide cluster in our new feature space is highly overfitted.
The results of SPF-EER and SV-EER therefore show a different tendency in the training and evaluation stages. We believe it is a reasonable solution to train end-to-end SASV systems on complete ASV and CM datasets to avoid the overfitting problem.

\begin{figure}[hbpt]
    \centering
    \includegraphics[width=8.5cm,height=4cm]{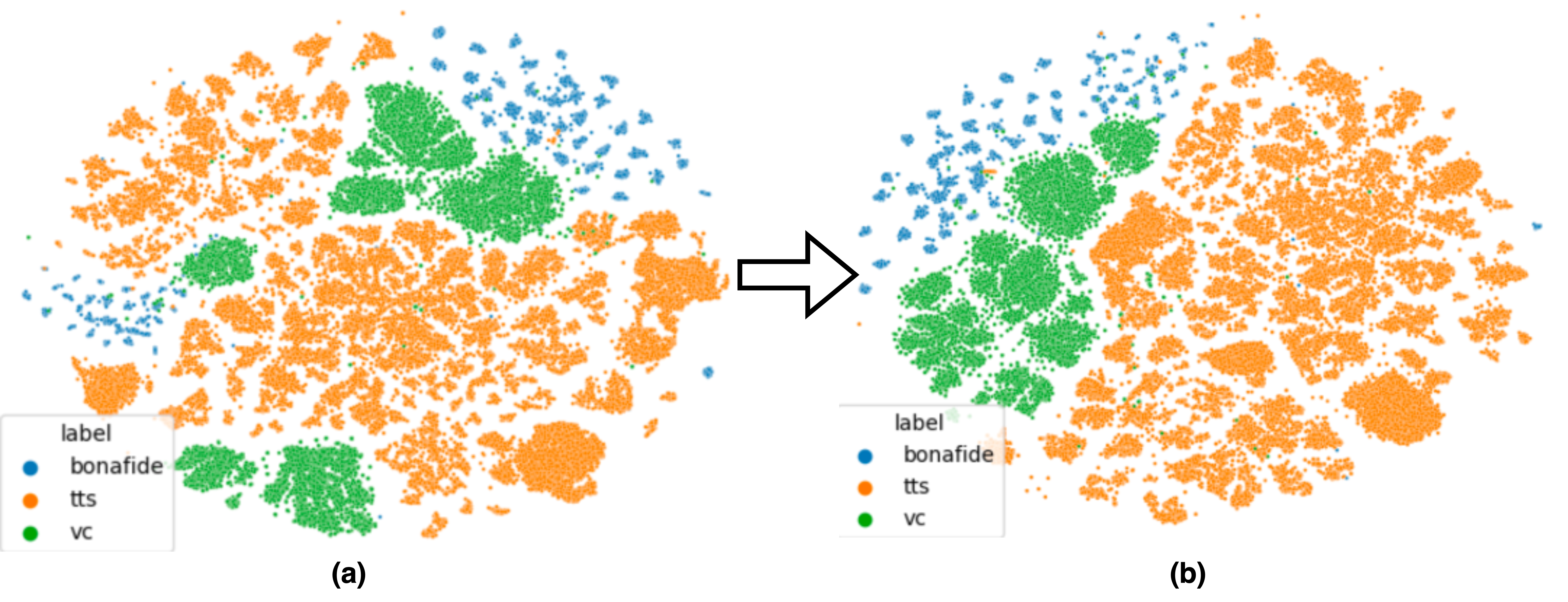}
    \vspace{-0.5cm}
    \caption{Visualization of the Feature Space in SA-SASV Using t-SNE.}
    \label{fig.vis}
    \vspace{-0.5cm}
\end{figure}

\textbf{Visualization}. To observe the updates of the features space produced by our encoder, we visualized utterances in the evaluation set using the t-SNE, as shown in Figure~\ref{fig.vis}. 
The left side (labeled (a)) shows the distribution of samples from the naive multi-task classifier without spoof-source-based triplet loss and the spoof aggregator. The right side (labeled (b)) shows the updated distribution using SA-SASV on the evaluation set. Compared to naive the multi-task classifier, both spoof attacks from TTS and VC tend to have denser clustering and cleaner boundaries, making TTS, VC, and bonafide easier to differentiate. 
\vspace{-0.2cm}
\subsubsection{Model Comparison with other SASV systems}
\vspace{-0.2cm}

We compared the characteristics and performance on the ASVSpoof 2019 LA dataset of SA-SASV with other SASV systems as shown in Table~\ref{table.comapre}. 
Compared to other ensemble-based systems, SA-SASV takes advantage of a single training phase, intending to build a single representation in the feature space for utterances from different sources.Our SA-SASV improves SASV-EER from 6.05\% (the prior best-performed INN(AUE) system) to 4.86\%.

\vspace{-0.1cm}
\section{Concluding Remarks}
\label{sec.conclusion}
We proposed an end-to-end SA-SASV model, which is optimized with multiple loss functions to aggregate TTS, VC, and different speakers separately. 
Results show that the feature space of SA-SASV is better able to distinguish spoof attacks and identify speakers versus prior published approaches.  Further, the SASV-EER is improved from the 6.05\% produced by prior state of the art approaches to 4.86\% without an ensembling strategy. A larger dataset and different encoders would likely boost the performance of the SV-EER and we will explore this in future work. The code described here is available in open-source form from: {\small\tt github.com/magnumresearchgroup/SA-SASV}. 
This paper is submitted to INTERSPEECH 2022.

%% file: tables/table_compare.tex
% \label{table.comapre}
% \caption{Comparison on characteristics and performance  of different  SASV systems.}
% \begin{table*}[hbt!]

% \usepackage{multirow}
\begin{table*}[hbt!]
\begin{tabular}{|c|c|c|cc|c|c|}
\hline
\multirow{2}{*}{Models}                                    & \multirow{2}{*}{Inputs}                                         & \multirow{2}{*}{Encoders}                                    & \multicolumn{2}{c|}{Training}                                                                             & \multirow{2}{*}{Ensemble} & \multirow{2}{*}{EER-SASV} \\ \cline{4-5}
                                                           &                                                                 &                                                              & \multicolumn{1}{c|}{Phase1}          & Phase2                                                             &                           &                           \\ \hline
SASV-Baseline1~\cite{jung2022sasv}                                             & \begin{tabular}[c]{@{}c@{}}raw waveforms,\\ Fbanks\end{tabular} & \begin{tabular}[c]{@{}c@{}}ECAPA-TDNN,\\ AASIST\end{tabular} & \multicolumn{1}{c|}{ASV, CM systems} & \textbackslash{}                                                   & Score                     & 19.15                     \\ \hline
SASV-Baseline2~\cite{jung2022sasv}                                             & \begin{tabular}[c]{@{}c@{}}raw waveforms,\\ Fbanks\end{tabular} & \begin{tabular}[c]{@{}c@{}}ECAPA-TDNN,\\ AASIST\end{tabular} & \multicolumn{1}{c|}{ASV, CM systems} & \begin{tabular}[c]{@{}c@{}}concatenated \\ embeddings\end{tabular} & Embeddings                & 8.76                      \\ \hline
\begin{tabular}[c]{@{}c@{}}Cascaded \\ CM/ASV~\cite{gomez2020joint}\end{tabular} & \begin{tabular}[c]{@{}c@{}}MFCC\\ STFT\end{tabular}             & \begin{tabular}[c]{@{}c@{}}LC-GRNN,\\ X-Vector\end{tabular}  & \multicolumn{1}{c|}{ASV, CM systems} & \textbackslash{}                                                   & \textbackslash{}          & 7.67                      \\ \hline
2-stage PLDA~\cite{gomez2020joint,sizov2015joint}                                               & MFCC                                                            & X-Vector                                                     & \multicolumn{1}{c|}{PLDA}            & PLDA                                                               & \textbackslash{}          & 28.40                     \\ \hline
Triplet TDNN~\cite{gomez2020joint,li2020joint}                                               & \begin{tabular}[c]{@{}c@{}}MFCC,\\ CQCC\end{tabular}            & TDNN                                                         & \multicolumn{1}{c|}{TDNN}            & \begin{tabular}[c]{@{}c@{}}PLDA(CM)\\ PLDA(ASV)\end{tabular}       & Score                     & 8.99                      \\ \hline
INN(AUE)~\cite{gomez2020joint}                                                   & \begin{tabular}[c]{@{}c@{}}MFCC,\\ STFT\end{tabular}            & \begin{tabular}[c]{@{}c@{}}LC-GRNN,\\ B-Vector\end{tabular}  & \multicolumn{1}{c|}{ASV, CM systems} & \begin{tabular}[c]{@{}c@{}}concatenated \\ embeddings\end{tabular} & Embeddings                & 6.05                      \\ \hline
\textbf{SA-SASV}                                           & \begin{tabular}[c]{@{}c@{}}raw waveforms,\\ Fbanks\end{tabular} & \begin{tabular}[c]{@{}c@{}}ECAPA-TDNN\\ ARawNet\end{tabular} & \multicolumn{1}{c|}{SA-SASV}         & \textbackslash{}                                                   & \textbackslash{}          & \textbf{4.86}             \\ \hline
\end{tabular}
\caption{Comparison on characteristics and performance  of different  SASV systems.}
\label{table.comapre}
\vspace{-0.7cm}
\end{table*}

%% file: tables/table_results.tex
% \usepackage{multirow}

% \usepackage{multirow}
\begin{table}[]
\begin{tabular}{|cc|c|c|c|}
\hline
\multicolumn{2}{|c|}{\multirow{2}{*}{Configuration}}                                                                 & \multirow{2}{*}{SASV} & \multirow{2}{*}{SV} & \multirow{2}{*}{SPF} \\
\multicolumn{2}{|c|}{}                                                                                               &                       &                     &                      \\ \hline
\multicolumn{2}{|c|}{ECAPA-TDNN}                                                                                     & 22.38                 & 0.83                & 29.32                \\ \hline
\multicolumn{2}{|c|}{SASV-Baseline1}                                                                                 & 19.15                 & 35.1                & 0.5                  \\ \hline
\multicolumn{2}{|c|}{SASV-Baseline2}                                                                                 & 8.75                  & 16.01               & 12.23                \\ \hline
\multicolumn{1}{|c|}{\multirow{4}{*}{Ours}} & SA-SASV                                                                & \textbf{4.86}         & 8.06                & \textbf{0.50}        \\ \cline{2-5} 
\multicolumn{1}{|c|}{}                      & w/o triplet                                                            & 5.82                  & 9.14                & 2.12                 \\ \cline{2-5} 
\multicolumn{1}{|c|}{}                      & \begin{tabular}[c]{@{}c@{}}w/o spoof\\ aggregator\end{tabular}         & 5.90                  & 9.96                & 0.68                 \\ \cline{2-5} 
\multicolumn{1}{|c|}{}                      & \begin{tabular}[c]{@{}c@{}}naive \\ multi-task classifier\end{tabular} & 5.58                  & 9.05      & 0.83                 \\ \hline
\end{tabular}
\caption{Ablation study on the AS-SASV system.}
\label{table.results}
\vspace{-0.8cm}
\end{table}